\documentclass[10pt,conference]{IEEEtran}
\usepackage{graphicx}
\usepackage{amsmath}
\usepackage{amsthm}
\usepackage{epsfig}
\usepackage{amsfonts}
\usepackage{amssymb}
\usepackage{epstopdf}

\newcommand{\argmax}{\mathop{\text{argmax}}}

\newcommand{\vw}{{\bf w}}

\newcommand{\vx}{{\bf x}}
\newcommand{\vy}{{\bf y}}
\newcommand{\vs}{{\bf s}}

\newcommand{\vb}{{\bf b}}
\newcommand{\va}{{\bf a}}

\newcommand{\ma}{{\bf A}}
\newcommand{\md}{{\bf D}}
\newcommand{\mh}{{\bf H}}
\newcommand{\mi}{{\bf I}}
\newcommand{\mc}{{\bf \Phi}}

\newcommand{\sap}{{\cal A}}

\newcommand{\gsm}{{\mathbb G}}
\newcommand{\sa}{{\mathbb A}}

\newcommand{\vmu}{{\boldsymbol \mu}}
\newcommand{\sn}{{\mathcal N}}

\newcommand{\E}{{\mathbb E}}

\newcommand{\Define}{\triangleq}

\begin{document}
\title{
{\huge On the Capacity and Performance of Generalized Spatial Modulation}
}

\author{T. Lakshmi Narasimhan$^\dagger$ and A. Chockalingam$^\ddagger$ \\
$\dagger$ Presently with National Instruments Private Limited,
Bangalore 560029  \\
$\ddagger$ Department of ECE, Indian Institute of Science, Bangalore 560012 
}

\maketitle

\begin{abstract}
Generalized spatial modulation (GSM) uses $N$ antenna elements but fewer
radio frequency (RF) chains ($R$) at the transmitter. Spatial modulation 
and spatial multiplexing are special cases of GSM with $R=1$ and $R=N$, 
respectively. In GSM, apart from conveying information bits through $R$ 
modulation symbols, information bits are also conveyed through the indices 
of the $R$ active transmit antennas. In this paper, we derive lower and 
upper bounds on the the capacity of a ($N,M,R$)-GSM MIMO system, where $M$ 
is the number of receive antennas. Further, we propose a computationally 
efficient GSM encoding (i.e., bits-to-signal mapping) method and a message 
passing based low-complexity detection algorithm suited for large-scale 
GSM-MIMO systems.
\end{abstract}


\section{Introduction}
\label{sec1}

Spatial modulation (SM) is emerging as a promising multi-antenna
modulation scheme (see \cite{sm1} and the references therein).
SM uses multiple antenna elements but only one radio frequency
(RF) chain at the transmitter. The single-RF chain feature of SM
brings in several advantages including reduced RF hardware complexity,
size and cost, no spatial interference, and simple detection at the
receiver, while offering the capacity and diversity benefits of
multi-antenna systems. In SM, only one antenna element is activated
in a given channel use and a QAM/PSK symbol is sent on the activated
antenna; the remaining antenna elements remain silent. The index of
the active antenna element also conveys information bits. It has been
shown that SM can achieve better performance than spatial multiplexing
(SMP) under certain conditions \cite{sm1}.

\begin{figure}[t]
\centering
\includegraphics[width=3.5in, height=2.6in]{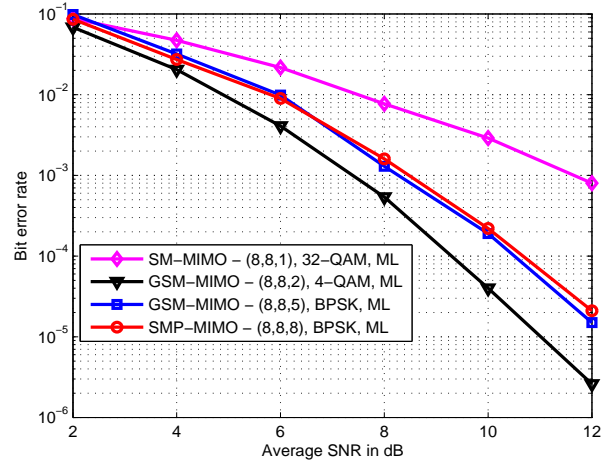}
\vspace{-6mm}
\caption{BER performance of different MIMO configurations with same
spectral efficiency of 8 bpcu.}
\vspace{-10mm}
\label{fig1}
\end{figure}

Generalized spatial modulation (GSM) is a generalization of SM, where
the transmitter uses multiple ($N$) antenna elements and more than one
($R$) RF chain \cite{gsm0c}. $R$ among the $N$ available
antenna elements are activated in a given channel use, and $R$ QAM/PSK
symbols are sent simultaneously on the active antennas. The indices of
the $R$ active antennas also convey information bits. Both SM and SMP
can be seen as special cases of GSM with $R=1$ and $R=N$, respectively.
It has been shown that for a given modulation alphabet and $N$, there 
exists an optimum $R\leq N$ that maximizes the spectral efficiency 
\cite{gsm1}. Further, for the same spectral efficiency, GSM can perform 
better than both SM and SMP \cite{sm3},\cite{sm4}. This is illustrated
in Fig. \ref{fig1} which shows the bit error rate (BER) performance of 
maximum likelihood (ML) decoding for different configurations of MIMO 
systems each with $N=M=8$ and same spectral efficiency of 8 bits per 
channel use (bpcu). We can see that the (8,8,2)-GSM MIMO system 
outperforms both the (8,8,1)-system (i.e., SM system) and (8,8,8-system 
(i.e., SMP system). This superior performance of GSM-MIMO motivates us 
to study its capacity limits. While most studies on SM/GSM in the 
literature so far have focused mainly on performance analysis and 
receiver algorithms, capacity of SM/GSM systems remains to be studied. 
The need for capacity analysis of SM has also been highlighted in 
\cite{sm1}. In \cite{chinacom}, the authors have obtained the capacity 
of SM for MISO systems through simulation. However, an analytical 
characterization of the capacity of SM/GSM has not been explored. 
Our contribution in this paper addresses this gap. In particular, we 
derive lower and upper bounds on the capacity of GSM. Another 
contribution is the proposal of low complexity encoding (i.e., 
bits-to-GSM signal mapping) and detection methods suited for GSM-MIMO 
systems with large number of antennas and RF chains. The proposed 
encoding makes use of combinadic representations in combinatorial number 
system, and the detection makes use of layered message passing.

The rest of the paper is organized as follows. Section \ref{sec2} 
introduces the GSM-MIMO system model. GSM-MIMO capacity bounds
are derived in Section \ref{sec3}. The proposed combinadics based 
GSM encoding scheme and the layered message passing based GSM 
detection scheme are presented in Section \ref{sec5}. Conclusions
are presented in Section \ref{sec6}.

\section{GSM-MIMO system model}
\label{sec2}
\begin{figure}[t]
\centering
\includegraphics[height=1.4in]{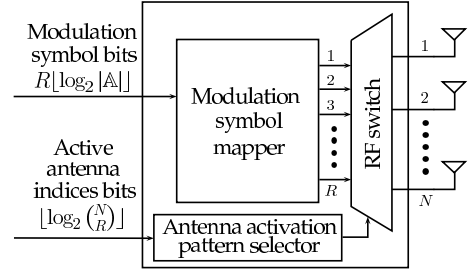}
\caption{GSM transmitter.}
\label{sysdia}
\vspace{-4mm}
\end{figure}

Consider a ($N,M,R$)-GSM MIMO system with $N$ antennas and $R$ RF chains
at the transmitter ($1\leq R \leq N$), and $M$ antennas at the receiver.
Figure \ref{sysdia} shows the GSM transmitter.
An $R\times N$ switch connects the $R$ RF chains to the $N$ transmit
antennas. In each channel use, $R$ out of $N$ transmit antennas are
chosen and activated. The remaining $N-R$ antennas remain silent.
The selection of the $R$ antennas to activate in a
channel use is done based solely on $\lfloor\log_2{N \choose R}\rfloor$
information bits (not based on CSIT). Therefore, the indices of the
active antennas convey $\lfloor\log_2{N \choose R}\rfloor$ information
bits per channel use. On the active antennas, $R$ modulation symbols 
(one on each active antenna) from a modulation alphabet $\sa$ are 
transmitted. This conveys $R\lfloor\log_2|\sa|\rfloor$ additional 
information bits. Therefore, the total number of bits conveyed in a 
channel use in GSM is given by
\begin{equation}
\eta_{\scriptsize{\mbox {gsm}}}=\underbrace{R\left\lfloor\log_2\sa\right\rfloor}_{\mbox{\scriptsize Modulation symbol bits}}+\underbrace{\bigg\lfloor\log_2{N \choose R}\bigg\rfloor}_{\mbox{\scriptsize Antenna index bits}} \quad \mbox{bpcu}.
\label{eta}
\end{equation}

\subsection{GSM signal set}
Let $\gsm$ denote the GSM signal set, which is the set of all possible
GSM signal vectors that can be transmitted. Therefore, if $\vx$ is
the $N\times1$ signal vector transmitted by the GSM transmitter, then
$\vx\in\gsm$. For example, for $N=4$, $R=2$, and BPSK, i.e.,
$\sa =\{\pm 1\}$, a GSM signal set can be as follows\footnote{A zero
in a coordinate in a GSM signal vector means the transmit antenna
corresponding to that coordinate is silent.}:

{\footnotesize
\begin{eqnarray}
\hspace{-0mm}
\gsm
\hspace{-3mm}&=&\hspace{-3mm}\left\{
\begin{bmatrix} +1 \\ +1 \\ 0 \\ 0\end{bmatrix},
\begin{bmatrix} +1 \\ -1 \\ 0 \\ 0\end{bmatrix},
\begin{bmatrix} -1 \\ -1 \\ 0 \\ 0\end{bmatrix},
\begin{bmatrix} -1 \\ +1 \\ 0 \\ 0\end{bmatrix},
\begin{bmatrix} +1 \\ 0 \\ +1 \\ 0\end{bmatrix},
\begin{bmatrix} +1 \\ 0 \\ -1 \\ 0\end{bmatrix},
\begin{bmatrix} -1 \\ 0 \\ -1 \\ 0\end{bmatrix},
\begin{bmatrix} -1 \\ 0 \\ +1 \\ 0\end{bmatrix}, \right.
\nonumber \\ & & \left.
\begin{bmatrix} +1 \\ 0  \\ 0\\ +1\end{bmatrix},
\begin{bmatrix} +1 \\ 0  \\ 0\\ -1\end{bmatrix},
\begin{bmatrix} -1 \\ 0  \\ 0\\ -1\end{bmatrix},
\begin{bmatrix} -1 \\ 0  \\ 0\\ +1\end{bmatrix},
\begin{bmatrix} 0\\ +1 \\ +1 \\ 0\end{bmatrix},
\begin{bmatrix} 0\\ +1 \\ -1 \\ 0\end{bmatrix},
\begin{bmatrix} 0\\ -1 \\ -1 \\ 0\end{bmatrix},
\begin{bmatrix} 0\\ -1 \\ +1 \\ 0\end{bmatrix}
\right \}. \nonumber
\end{eqnarray}
}

Let $\vs$ denote the $R\times1$ vector of modulation symbols that are
to be transmitted in a channel use over the $R$ chosen antennas, 
i.e., $\vs\in\sa^R$.
 
{\bf Definition:} Define a matrix $\ma$ of size $N\times R$ as the 
{\em antenna activation pattern matrix}. The matrix $\ma$ represents 
a particular choice of $R$ antennas from the available $N$ antennas, 
such that the $N\times1$ GSM signal vector $\vx=\ma\vs \in \gsm$. The 
matrix $\ma$ is a sparse matrix consisting only of $1$'s and $0$'s, 
with exactly one non-zero entry in every column and one non-zero entry 
in $R$ rows. If $I_1,I_2,\cdots,I_r,\cdots,I_R$ are the indices of 
the chosen antennas, then $\ma$ is constructed as
\begin{equation}
A_{ij}=\left\{\begin{array}{ll}
1 & \mbox{if } j=r \mbox{ and } i=I_r \\
0 &  \mbox{otherwise},
\end{array} \right.
\label{aap}
\end{equation} 
where $A_{ij}$ denotes the element in the $i$th row and $j$th column
of $\ma$. 

For example, in a system with $N=8$ and $R=4$, to activate 
antennas 1, 3, 6 and 8, the matrix $\ma$ is given by
\begin{equation}
\ma \ = \ 
\left[
\begin{array}{cccc}
1 &0 &0 &0\\
0 &0 &0 &0\\
0 &1 &0 &0\\
0 &0 &0 &0\\
0 &0 &0 &0\\
0 &0 &1 &0\\
0 &0 &0 &0\\
0 &0 &0 &1\\
\end{array}\right].
\label{aap1}
\end{equation}
Note that the indices of the non-zero rows in matrix $\ma$ give the 
support of the GSM signal vector $\vx$. Out of the $N\choose R$ possible 
antenna activation choices, only $2^{\lfloor \log_2{N\choose R}\rfloor}$ 
are needed for signaling. Let $\sap$ denote this set of all allowed 
antenna activation pattern matrices, where 
$|\sap|=2^{\lfloor\log_2{N \choose R}\rfloor}$ and $\ma \in \sap$. 
Let $L\Define|\sap|$. Now, $\gsm$ is given by
\begin{eqnarray}
\gsm =\big\{\vx : \vx=\ma\vs, \mbox{ for some } \ma\in\sap, \vs\in\sa^R\big \}.
\label{gsmset}
\end{eqnarray}

Note that any GSM signal vector $\vx_j\in\gsm$, $j=1,\cdots,|\gsm|$,
can be represented as $\vx_j=\ma_i\vs_k$ with some $\ma_i\in\sap$,
$i=1,\cdots,|\sap|$ and $\vs_k\in\sa^R$,  $k=1,\cdots,|\sa|^R$, and
$|\gsm|=|\sap||\sa|^R$. Conversely, given any $\ma_i\in\sap$ and
$\vs_k\in\sa^R$, there exists a GSM signal vector $\vx_j\in\gsm$ such 
that $\vx_j=\ma_i\vs_k$. Since $\ma_i$ and $\vs_k$ are chosen by two 
independent information bit sequences, $\ma_i$ and $\vs_k$ are 
independent. That is, $p(\vx)=p(\ma\vs)=p(\ma)p(\vs)$.
Thus, the above system model of GSM-MIMO decouples the choice of active 
antennas and the transmitted modulation symbols into two independent random 
variables, namely $\ma$ and $\vs$. This makes the GSM-MIMO system analysis 
simpler.

The $M\times 1$ received signal vector $\vy=[y_1 \ y_2 \cdots \ y_{M}]^T$
at the receiver can be written as
\begin{eqnarray}
\vy& = & \mh\vx+\vw \ = \ \mh\ma\vs+\vw,
\label{sysmodel}
\end{eqnarray}
where $\vx \in \gsm$ is the transmit vector, $\mh \in \mathbb{C}^{M\times N}$
is the channel gain matrix whose 
$(i,j)$th entry $H_{i,j}\sim \mathcal{CN}(0,1)$ 
denotes the complex channel gain from the $j$th transmit antenna to the 
$i$th receive antenna, and $\vw = [w_1 \ w_2 \cdots \ w_{M}]^T$ is the 
noise vector whose entries are modeled as complex Gaussian with zero mean 
and variance $\sigma^2$. Since $\vx$ is obtained based only on 
$\eta_{\scriptsize{\mbox{gsm}}}$ 
information bits, $\vx$ and $\mh$ are independent. Since $\vx=\ma\vs$,
$\ma\vs$ and $\mh$ are independent. Also, ${\bf A}$ and ${\bf s}$ are 
independent. So, $\ma$ and $\mh$ are independent. So, the mutual 
information between $\vx$ and $\vy$ in GSM is given by 
$I(\vx;\vy|\mh)=I(\ma,\vs;\vy|\mh)$.

\section{GSM-MIMO capacity bounds}
\label{sec3}
The capacity of a spatially multiplexed $M\times N$ MIMO channel 
with channel state information at the receiver is \cite{tse}
\begin{equation}
C=\E_{\mh}\Big\{ \log_2 \big[ \det (\mi_{M} + \frac{1}{\sigma^2}\mh\mc\mh^H) \big] \Big\},
\label{cap}
\end{equation}
where $\mc$ is the covariance matrix of the transmit signal vector
with elements from Gaussian codebook. For i.i.d. modulation symbols
and total power $\sigma^2_x$, the capacity expression becomes 
\begin{equation}
C=\E_{\mh}\Big\{ \log_2 \big[ \det (\mi_{M} + \frac{\sigma^2_x}{N\sigma^2}\mh\mh^H) \big] \Big\}.
\label{capsmp}
\end{equation}
Here, we are interested in the capacity of GSM-MIMO, where the symbols 
sent on the active antennas are from Gaussian codebook. The capacity of
GSM-MIMO can be written as
\begin{eqnarray}
C_{GSM}&=& \E_{\mh}(I(\vx; \vy))\nonumber\\
&=&\E_{\mh}(h(\vy)-h(\vy|\vx))=\E_{\mh}(h(\vy)-h(\vw))\nonumber\\
&=&\E_{\mh}(h(\vy)-\log_2[\det (\pi e\sigma^2\mi_{M} )]),
\label{gsmcap}
\end{eqnarray} 
where $h(.)$ denotes the differential entropy. To compute $C_{GSM}$, we need
to evaluate $h(\vy)$, which requires the knowledge of the distribution of $\vy$.
From (\ref{sysmodel}), we can see that the distribution of $\vy$ is a 
Gaussian mixture given by
\begin{eqnarray}
p(\vy)&=&\sum_{i=1}^{L}p(\vy,\ma_i)=\sum_{i=1}^{L}p(\vy|\ma_i)p(\ma_i)\nonumber\\
&=&\sum_{i=1}^{L}\sn(\vmu_i,\mc_i)p_i,
\label{ydist}
\end{eqnarray}
where 
\begin{eqnarray}
p_i&=&p(\ma_i), \quad\quad \ma_i\in\sap, \quad\quad i=1,2,\cdots,L,\\
\vmu_i&=&\E(\vy|\ma_i)\nonumber\\&=&\E(\mh\ma_i\vs+\vw|\ma_i)={\bf 0},\\
\mc_i&=&\E(\vy\vy^H|\ma_i)=\E[(\mh\ma_i\vs+\vw)(\mh\ma_i\vs+\vw)^H|\ma_i]\nonumber\\
&=&\mh\ma_i\E_{\vs}(\vs\vs^H)\ma_i^H\mh^H + \sigma^2\mi_{M}.
\label{covar}
\end{eqnarray}
If all the possible antenna activation patterns are equally likely, then
$p_i=\frac{1}{L}$. When the transmitted modulation symbols are independent 
of each other, $\E(\vs\vs^H)=\frac{\sigma^2_x}{R}\mi_{R}$, 
Now, (\ref{ydist}) becomes
\begin{equation}
\vy\sim\frac{1}{L}\sum_{i=1}^{L}\sn\Big({\bf 0},\frac{\sigma^2_x}{R}\mh\ma_i\ma_i^H\mh^H + \sigma^2\mi_{M}\Big).
\end{equation} 
The differential entropy of $\vy$ is given by
\begin{equation}
h(\vy)=-\frac{1}{L}\sum_{i=1}^{L}\int_{\vy}\sn({\bf 0},\mc_i)\log_2\Big(\frac{1}{L}\sum_{i=1}^L\sn({\bf 0},\mc_i)\Big) d\vy.
\end{equation}
It is difficult to get a closed-form solution to the above expression. Hence,
we try to bound it above and below by the following techniques.

{\em Lower bound 1, $L_1$}:
Since $-\log(.)$ is a convex function, by Jensen's inequality, 
\[
\E[-\log p(y)]\geq-\log\E[p(y)].
\]
Hence, the differential entropy $h(\vy)$ can be lower 
bounded as
\begin{eqnarray}
h(\vy)&=&\E[-\log_2 p(\vy)]\nonumber \\
&\geq&-\log_2\int p^2(\vy)d\vy \nonumber \\
&=&-\log_2\int \Big(\frac{1}{L}\sum_{i=1}^L\sn({\bf 0},\mc_i)\Big)^2 d\vy \nonumber \\
&=&-\log_2\bigg\{ \frac{1}{L^2\pi^{M}} \sum_{i=1}^L\sum_{j=1}^L\frac{\det((\mc_i^{-1}+\mc_j^{-1})^{-1})}{\det(\mc_i)\det(\mc_j)}\nonumber \\
&&\hspace{2cm}\underbrace{\int \frac{e^{-\vy^H(\mc_i^{-1}+\mc_j^{-1})\vy}}{\pi^{M}\det((\mc_i^{-1}+\mc_j^{-1})^{-1})} d\vy}_{= 1} \bigg\}\nonumber \\
&=&-\log_2\bigg\{ \frac{1}{L^2\pi^{M}} \sum_{i=1}^L\sum_{j=1}^L\frac{1}{\det(\mc_i+\mc_j)} \bigg\} \label{l1}\\
&\Define& l_1.\nonumber
\end{eqnarray}
From (\ref{gsmcap}) and (\ref{l1}), a lower bound on $C_{GSM}$ can be obtained as
\begin{equation}
C_{GSM}\geq L_1\Define \E_{\mh}(l_1-\log_2 \det (\pi e\sigma^2\mi_{M} )).
\label{lb1}
\end{equation}

{\em Lower bound 2, $L_2$}:
Since differential entropy is a concave function, we can write
\begin{eqnarray}
h(\vy)&\hspace{-3mm}=&\hspace{-3mm}h\Big(\frac{1}{L}\sum_{i=1}^L\sn({\bf 0},\mc_i)\Big) \nonumber\\
&\hspace{-3mm}\geq&\hspace{-3mm}\frac{1}{L}\sum_{i=1}^Lh\big(\sn({\bf 0},\mc_i)\big) 
=\frac{1}{L}\sum_{i=1}^L\log_2 \det \big(\pi e \mc_i) \Define l_2.\nonumber
\label{lb2d}
\end{eqnarray}
From the above equation, $C_{GSM}$ can be lower bounded as
\begin{equation}
C_{GSM}\geq L_2\Define \E_{\mh}(l_2-\log_2 \det (\pi e\sigma^2\mi_{M} )).
\label{lb2}
\end{equation}

Based on the two lower bounds $L_1$ and $L_2$, a refined lower bound on 
GSM-MIMO capacity is given by
\begin{equation}
C_{GSM}\geq L \Define \text{max}(L_1,L_2).
\label{lb}
\end{equation}

{\em Upper bound 1, $U_1$}:
By the property of entropy, we can write
\begin{eqnarray}
h(\vy, \ma)&=&-\sum_{\forall \ma} \int p(\vy,\ma)\log_2p(\vy,\ma) d\vy \nonumber \\
&=& -\sum_{\forall \ma}\int p(\vy,\ma)\log_2p(\vy|\ma)d\vy \nonumber \\
&& -\sum_{\forall \ma}\log_2p(\ma)\int p(\vy,\ma) d\vy \nonumber\\
&=& h(\vy|\ma)+H(\ma).\nonumber
\end{eqnarray}
Further, we also have
\begin{eqnarray}
h(\vy)&=&h(\vy, \ma)-h(\ma|\vy) \leq h(\vy, \ma). \nonumber
\end{eqnarray}
Using the above properties, an upper bound for $h(\vy)$ can be written as
\begin{eqnarray}
h(\vy)&\leq&h(\vy, \ma)=h(\vy|\ma)+H(\ma) \nonumber \\
&=&\int\sum_{\forall \ma} p(\vy,\ma)\log_2\frac{p(\ma)}{p(\vy,\ma)} d\vy+ \sum_{i=1}^L-p_i\log_2 p_i \nonumber \\
&=&\frac{1}{L}\sum_{i=1}^{L} \log_2 \det \big(\pi e \mc_i)+ \log_2 L \Define u_1.\nonumber
\label{upb1}
\end{eqnarray}
Now, an upper bound on $C_{GSM}$ can be written as 
\begin{equation}
C_{GSM}\leq U_1\Define \E_{\mh}(u_1-\log_2 \det (\pi e\sigma^2\mi_{M} )).
\label{ub1}
\end{equation}

{\em Upper bound 2, $U_2$}:
Here, we approximate the probability distribution of $\vy$ to a Gaussian
distribution. This leads to an upper bound because the entropy of any 
random variable is bounded above by the entropy of a Gaussian random variable 
with the same mean and variance. The mean and covariance 
of $\vy$ are given by
\begin{eqnarray}
\E(\vy)&=&{\bf 0}, \nonumber \\
\E(\vy\vy^H)&=& \mh\E_{\ma}[\ma_i\E_{\vs}(\vs\vs^H)\ma_i^H]\mh^H + \sigma^2\mi_{M}\nonumber \\
&=& \mh\bigg(\frac{1}{L}\sum_{i=1}^L\ma_i\Big(\frac{\sigma^2_x}{R}\mi_{R}\Big)\ma_i^H\bigg)\mh^H + \sigma^2\mi_{M}\nonumber \\
&=& \frac{\sigma^2_x}{RL}\mh\Big(\sum_{i=1}^L\md_i\Big)\mh^H + \sigma^2\mi_{M},
\label{upb1d1}
\end{eqnarray}
where $\md_i\Define\ma_i\ma^H_i$.
Let $\{I_1^i, I_2^i, \cdots, I_r^i, \cdots, I_{R}^i\}$ be the set of 
active antenna indices that corresponds to the antenna activation pattern
matrix $\ma_i$. It can then be seen that $\md_i$ is a diagonal 
matrix, such that
\[
(D_i)_{j,k}=\left\{\begin{array}{ll}
1 & \mbox{if } j=k=I\in \{I_1^i, I_2^i, \cdots, I_{R}^i\}\\
0 & \mbox{otherwise},
\end{array} \right.
\]
where $(D_i)_{j,k}$ is the element in the $j$th row and $k$th column
of $\md_i$. Assuming that all $N \choose R$ activation patterns 
are allowed, i.e., $|\sap|=L={N \choose R}$, the number of times 
any particular antenna will be active among the $N \choose R$ 
activation patterns is $N-1 \choose R-1$. Therefore, 
$\md={N-1 \choose R-1}\mi_{N}$ and (\ref{upb1d1}) becomes
\begin{eqnarray}
\E(\vy\vy^H)
&=&\frac{\sigma^2_x}{N}\mh\mh^H + \sigma^2\mi_{M} \Define \Phi'.\nonumber
\end{eqnarray}
Now, an upper bound on the GSM-MIMO capacity is given by
\begin{eqnarray}
C_{GSM}&\leq&U_2\Define \E_{\mh}(\log_2\det (\pi e\Phi' )-\log_2 \det (\pi e\sigma^2\mi_{M} ))\nonumber \\
&=&\E_{\mh}\Big\{ \log_2 \big[ \det (\mi_{M} + \frac{\sigma^2_x}{N\sigma^2}\mh\mh^H) \big] \Big\},
\label{ub2}
\end{eqnarray}
which is the same as the capacity of a $M\times N$ spatially multiplexed
MIMO system.

Based on the two upper bounds $U_1$ and $U_2$, a refined upper bound on 
GSM capacity can be obtained as
\begin{equation}
C_{GSM}\leq U \Define \text{min}(U_1,U_2).
\label{ub}
\end{equation}

\begin{figure}[t]
\centering
\includegraphics[width=3.5in,height=2.6in]{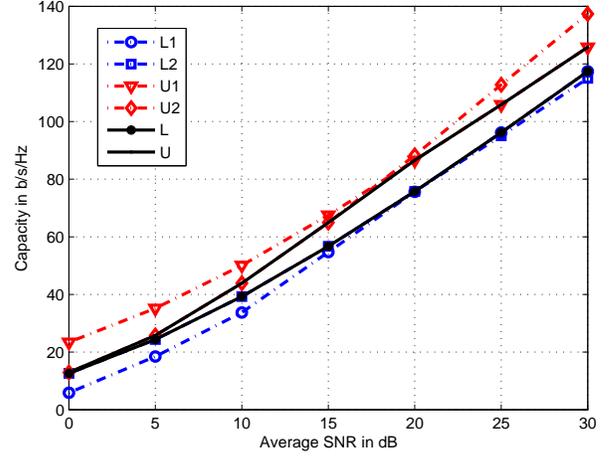}
\vspace{-6mm}
\caption{Capacity bounds for GSM-MIMO system with $N=16,R=12$, and $M=16.$}
\label{capfig1}
\vspace{-4mm}
\end{figure}

\begin{figure}[t]
\centering
\includegraphics[width=3.5in,height=2.75in]{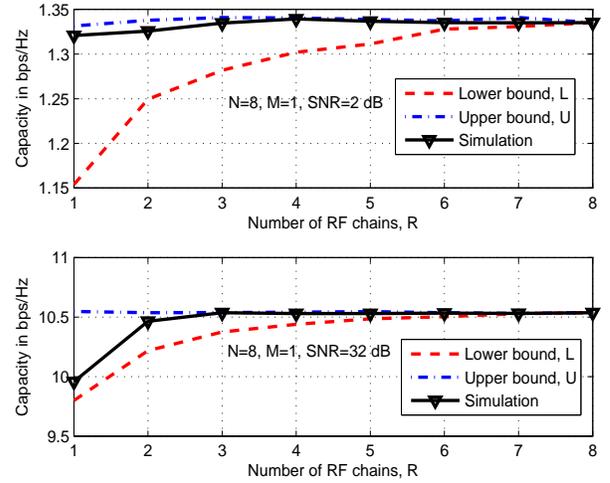}
\vspace{-8mm}
\caption{Comparison of simulated capacity and bounds for GSM-MIMO system 
with $N=8$, $M=1$, 
and varying $R$.}
\vspace{-6mm}
\label{capfig2}
\end{figure}

{\em Numerical results}:
We evaluated the lower and upper bounds on the GSM-MIMO capacity
for different system configurations. Figure \ref{capfig1} shows the
lower and upper bounds for GSM-MIMO systems with $N=16$, $R=12$, $M=16$.
We see that the lower bound $L_2$ and upper bound $U_2$ are tighter at
low SNRs. Whereas, the lower bound $L_1$ and upper bound $U_1$ are
tighter at high SNRs. It can be noted from the figure that the lower
and upper bounds are very close at low SNRs; so, in this regime, the
GSM-MIMO capacity is almost same as that of a spatially multiplexed
MIMO system with the same $N$ and $M$. In Fig. \ref{capfig2}, we compare
the bounds on GSM-MIMO capacity with the true GSM-MIMO capacity obtained
through simulation. We consider GSM-MIMO with $N=8$, $M=1$, SNR $=2,32$ dB,
and varying $R$. In the top figure of Fig. \ref{capfig2}, we observe that, 
at SNR = 2 dB, the gap between the upper and lower bounds is $U-L=0.188$ 
bps/Hz for $R=1$ and $U-L<10^{-2}$ bps/Hz for $R>5$. Also, in the bottom
figure of Fig. \ref{capfig2}, at SNR = 32 dB, the gap is $U-L=0.782$ bps/Hz 
for $R=1$ and $U-L<10^{-2}$ bps/Hz for $R>6$.

\section{Low-complexity GSM-MIMO encoding/decoding}
\label{sec5}
It is known that large-scale MIMO systems provide a variety of 
advantages \cite{book}. However, for large values of $N,M,R$ 
in GSM-MIMO, the encoding and decoding 
complexities become prohibitively large. In this section, we propose 
low-complexity methods to encode and detect GSM signals that scale
well for large $N,M,R$.

\subsection{GSM encoding using combinadics}
\label{subsec5a}
In GSM-MIMO, $\lfloor\log_2{N \choose R}\rfloor$ bits are used to choose
an activation pattern matrix $\ma$ from $\sap$, and 
$R\left\lfloor\log_2\sa\right\rfloor$ bits are used to generate $\vs$ 
from $\sa^R$. While the mapping of $R\left\lfloor\log_2\sa\right\rfloor$ 
bits to modulation symbols in $\vs$ is straight-forward, the mapping of 
$\lfloor\log_2{N \choose R}\rfloor$ bits to a choice of activation 
pattern is not. A table or a map of bit sequence to activation patterns 
has to be maintained both at the transmitter and receiver. For large
values of $N,M,R$, the size of this map can become prohibitively large. 
For example, if $N=64, R=32$, then 
{\small
$|\sap|={64 \choose 32}\approx1.83\times10^{18}\approx2^{60}$.
}
Implementation of an encoding map of this size is impractical. To 
alleviate this problem, we use {\em combinadic} representations 
in combinatorial number system.

{\bf Definition}: The combinadic of a number $n\in[0,{N \choose R}-1]$ 
is the $R$-tuple $(N_1, N_2, \cdots, M)$ such that 
$n=\sum_{i=1}^R{N_i\choose i}$ and $N_1<N_2<\cdots<M<N$.

The values of $N_i$ for a given $n$ can be obtained as \cite{combinadic} 
\[
N_i=\mbox{Largest non-negative integer s.t. } n-\sum_{j=i}^R {N_j \choose j} \geq 0.
\]

The following encoding procedure maps the bits to antenna activation
pattern in GSM. Let $\eta_a\Define\lfloor\log_2{N \choose R}\rfloor$.
\begin{enumerate}
\item Accumulate $\eta_a$ bits to form the bit sequence
$\vb=[b_{\eta_a-1},\cdots,b_1,b_0]$.
Obtain {\small $g(\vb)=\sum_{i=0}^{\eta_a-1} 2^ib_i$}.
\item Find the combinadic of $g(\vb)$.
\item Construct $\ma$ matrix such that the indices of the $R$
non-zero rows of $\ma$ are given by the combinadic of $g(\vb)$.
\end{enumerate}
An example of the encoding map for $N=4, R=3$ based of the above
procedure is illustrated in Table \ref{tab1}. The demapping procedure 
is listed below.
\begin{enumerate}
\item Obtain the combinadic corresponding to the decoded antenna 
activation pattern matrix.
\item Compute $g(\hat{\vb})$ from this combinadic as 
$g(\hat{\vb})=\sum\limits_{i=1}^R{N_i\choose i}$.
\item Demap $g(\hat{\vb})$ to get $\hat{\vb}$.
\end{enumerate}
The above low complexity mapping/demapping procedures allow practical 
implementation of GSM encoding for large $N,M,R$.
\begin{table}
\centering
\begin{tabular}{|c|c|c|c|}
\hline
$\vb$ & $g(\vb)$ & combinadic of $g(\vb)$  & Active antenna\\
 &  & $(N_1, N_2, N_3)$ & indices\\ \hline\hline
  $00$ & $0$ & $(0,1,2)$ & $(1,2,3)$ \\ \hline
  $01$ & $1$ & $(0,1,3)$ & $(1,2,4)$ \\ \hline
  $10$ & $2$ & $(0,2,3)$ & $(1,3,4)$ \\ \hline
  $11$ & $3$ & $(1,2,3)$ & $(2,3,4)$ \\ \hline
\end{tabular}
\vspace{4mm}
\caption{\label{tab1} Encoding map for GSM system with $N=4, R=3$. }
\vspace{-8mm}
\end{table}

{\em Example:} 
Let $n_t=10, n_{rf}=4$. Then, $|\sap|=2^7=128$.\\
{\em Mapping}: Let $\vb=[0 0 1 0 0 1 1]$ be the bit sequence that is to 
be transmitted. The decimal equivalent of $\vb$ is $d(\vb)=19$, and the 
combinadic of $19$ is $(N_1,N_2,N_3,N_4)=(0,1,4,6)$. The corresponding 
indices of the activated antennas are $(1,2,5,7)$.\\
{\em Demapping}: Let $(3,4,5,6)$ be the detected antenna activity pattern 
at the receiver. The combinadic of this antenna activity pattern is 
$(2,3,4,5)$, and the decimal equivalent is $14$. The bit sequence for 
this decimal value of 14 is $[0 0 0 1 1 1 0]$. This gives the corresponding 
bit sequence of the detected antenna activity pattern at the receiver.

\subsection{GSM Detection: Layered message passing algorithm}
\label{subsec5b}
The maximum a posteriori  probability (MAP) detection rule for GSM-MIMO 
is given by
\begin{eqnarray}
\label{map}
\hat{\vx} = \argmax_{\vx\in \gsm} \ p(\vx\mid\vy). 
\end{eqnarray}
Note that, $|\gsm|=2^{\eta_{\scriptsize{\mbox{gsm}}}}$. Therefore, the 
exact computation of 
(\ref{map}) requires exponential complexity in $N, R$. To address this 
problem, here we propose a low complexity layered message passing (LaMP)
algorithm which gives an approximate solution to (\ref{map}). In the 
proposed LaMP algorithm, we define four sets of nodes and exchange 
messages between them in layers. There are two layers of message 
exchanges, one corresponds to the detection of modulation symbols
and the other corresponds to the detection of the antenna activation 
pattern. In order to describe this algorithm, we define a new variable 
as follows.

{\bf Definition}: A variable $a_i$ is called the antenna activity 
indicator if $a_i=1$ if the $i$th antenna is active, else $a_i=0$. 
That is, $a_i={\bf 1}_{\{i\text{th antenna is active}\}}$, where 
${\bf 1}$ is the indicator function. 

\begin{figure}[h]
\centering
\includegraphics[height=1.8in]{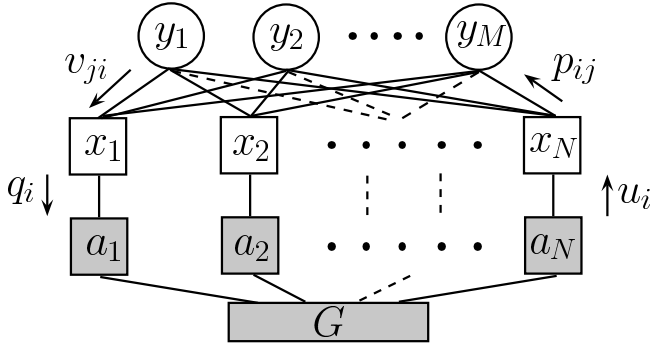}
\caption{Graphical model and messages passed in the proposed LaMP detector.}
\label{graph}
\vspace{-4mm}
\end{figure}

Therefore, $x_i=a_is,s\in\sa$. Note that $\sum_{i=1}^{N}a_i=R$, which 
we call as the GSM system constraint $G$. Let 
$\va\Define[a_1,a_2,\cdots,a_{N}]$. Now, $p(\vx\mid\vy)$ in (\ref{map}) 
can be written as
\begin{eqnarray}
p(\vx,\va|\vy)&\propto&p(\vy|\vx,\va)p(\vx,\va)\nonumber\\
&=&p(\vy|\vx)p(\vx|\va)p(\va)\nonumber\\
&=&\Big\{\prod_{j=1}^{M} p(y_i|\vx) \prod_{i=1}^{N}p(x_i|a_i)\Big\}p(\va).
\label{probs}
\end{eqnarray}
Thus, by defining a new layer of variables $a_i$s corresponding to antenna 
activity, we have effectively decoupled the dependences present among the 
elements of the transmit vector $\vx$. Based on (\ref{probs}), we model 
the GSM-MIMO system as a graph with four types of nodes, namely, 
\begin{itemize}
\item $M$ observation nodes corresponding to the elements of $\vy$, 
\item $N$ variable nodes corresponding to the elements of $\vx$, 
\item $N$ antenna activity nodes corresponding to the elements of $\va$, and 
\item a constraint node corresponding to the GSM constraint $G$. 
\end{itemize}
This is illustrated in Fig. \ref{graph}. On this graph, we 
iteratively pass messages between the nodes and obtain the marginal 
probabilities of the transmitted symbols. 

The different messages passed in this graph are 
\begin{enumerate}
\item
$v_{ji}$: from observation node $y_j$ to variable node $x_i$,
\item
$p_{ij}$: from variable node $x_i$ to observation node $y_j$, 
\item
$q_{i}$: from variable node $x_i$ to antenna activity node $a_i$, and 
\item 
$u_{i}$: from antenna activity node $a_i$ to variable node $x_i$. 
\end{enumerate}
The messages are exchanged between two layers, namely,
\begin{enumerate}
\item {\em Layer 1}: observation nodes and variable nodes (denoted by 
unshaded nodes in
Fig \ref{graph}); this layer generates approximate a posteriori 
probabilities of the individual elements of $\vx$, and
\item {\em Layer 2}: antenna activity nodes and GSM constraint node 
(denoted by shaded nodes in Fig \ref{graph}); this layer generates 
approximate a posteriori probabilities of the individual elements of $\va$.
\end{enumerate}
In constructing the messages at the variable nodes, we employ a 
Gaussian approximation of the interference as described below. This 
significantly reduces the detection complexity. From (\ref{sysmodel}), 
we can write
\begin{eqnarray}
y_j=H_{ji}x_i + g_{ji},\quad g_{ji}\Define\sum_{l=1\atop{l\neq i}}^N H_{jl}x_l+w_j.
\end{eqnarray}
where $i=1,2,\cdots,N$ and $j=1,2,\cdots,M$. We approximate $g_{ji}$ to 
be Gaussian. Then,
\begin{eqnarray}
\mu_{ji}&\Define&\E(g_{ji}) \nonumber \\
& = & \sum_{l\neq i}^{} H_{jl}\E(x_l) \nonumber \\
& = & \sum_{l\neq i}^{} H_{jl} \sum_{x\in\sa\cup0}xp_{ij}(x), 
\end{eqnarray}
and
\begin{eqnarray}
\sigma_{ji}^2&\Define&\mbox{Var}(g_{ji}) \nonumber \\
& = & \sigma^2+\sum_{l\neq i}^{} H_{jl}^2\mbox{Var}(x_l) \nonumber \\
& = & \sum_{l\neq i}^{} H_{jl}^2\hspace{-1mm}\sum_{x\in\sa\cup0}\hspace{-1mm}x^2p_{ij}(x). 
\end{eqnarray}
Using the above approximation, we derive the messages as follows.

{\em Layer 1}:
\begin{eqnarray}
v_{ji}(x)&\Define&\Pr(x_i=x|y_j) \nonumber \\
& \approx & \frac{1}{\sigma_{ji}^2\sqrt{2\pi}}\exp\Big(\frac{-(y_j-\mu_{ji}-H_{ji}x)^2}{2\sigma_{ji}^2}\Big),
\end{eqnarray}
The APP of the individual elements of $\vx$ is given by
\begin{eqnarray}
p_{ij}(x) & \Define & \Pr(x_i=x|\vy_{\setminus j}) \nonumber \\
& \approx & \prod_{k=1\atop{k\neq j}}^M\Pr(x_i=x|y_k) \nonumber \\
& \propto & u_i(x^{\odot})\hspace{-0.0mm}\prod_{k\neq j}^{}\hspace{-0mm}v_{ki}(x),
\end{eqnarray}
where $\vy_{\setminus j}$ denotes the set of all elements of $\vy$ 
except $y_j$, and
\[
x^{\odot} = \left\{\begin{array}{ll}
0 & \mbox{if } x=0\\
1 & \mbox{else}. \end{array}\right.
\]
The inclusion of $u_i(.)$ in the computation of this APP helps us to relieve
the elements of $\vx$ from the dependencies on the antenna activity pattern.
 
{\em Layer 2}:
The APP estimate of $a_i$ from the variable nodes is
\vspace{-3mm}
\begin{eqnarray}
q_{i}(b) & \Define& \Pr(a_i=b|\vx) \nonumber \\
& \approx &
\left\{ \def\arraystretch{1.4} \begin{array}{ll}
\sum\limits_{x\in\sa}\prod\limits_{k=1}^M\Pr(x_i=x|y_k), & \mbox{if } b=1 \\
\prod\limits_{k=1}^M\Pr(x_i=0|y_k), &  \mbox{if } b=0
\end{array} \right.\nonumber\\
&\propto &\left\{ \def\arraystretch{1.4}\begin{array}{ll}
\sum\limits_{x\in\sa}\prod\limits_{k=1}^M v_{ki}(x) & \mbox{if } b=1 \\
\prod\limits_{k=1}^Mv_{ki}(0), &  \mbox{if } b=0,
\end{array} \right.
\end{eqnarray}
The APP estimate of $a_i$ after processing GSM constraint $G$ is
\begin{eqnarray}
u_{i}(b)&\Define&\Pr(a_i=b|\vx_{\setminus i}) \nonumber \\
& \propto & \left\{\begin{array}{ll}
\Pr(\sum\limits_{l\neq i}^{} a_l=R-1|\va_{\setminus i}), & \mbox{if } b=1 \\
\Pr(\sum\limits_{l\neq i}^{}a_l=R|\va_{\setminus i}), &  \mbox{if } b=0
\end{array} \right. \nonumber \\
&\approx &\left\{\begin{array}{ll} 
\phi_i(R-1) & \mbox{if } b=1 \\
\phi_i(R) &  \mbox{if } b=0,
\end{array} \right.
\end{eqnarray}
where  
$\Pr(\sum\limits_{l=1,l\neq i}^{N} a_l=R-1|\va_{\setminus i})$ denotes 
the probability that the antenna activity pattern satisfies the GSM 
signal constraint $G$ (i.e., $\sum_{i=1}^{N}a_i=R$) given that the 
$i$th antenna is active, and
$\Pr(\sum\limits_{l=1,l\neq i}^{N}a_l=R|\va_{\setminus i})$ denotes 
the probability that the antenna activity pattern satisfies $G$ given 
that the $i$th antenna is not active. Since this probability involves 
the summation of $N-1$ random variables, it is evaluated as 
$\phi_i=\bigotimes\limits_{l=1\atop{l\neq i}}^{N} q_l$, where 
$\bigotimes$ is the convolution operation, and $\phi_i(.)$ is a 
probability mass function with probability masses at $N$ points 
($0,1,2,\cdots,N-1$). 

{\em Message passing}:
The schedule of the messages passed in the LaMP algorithm is as follows:
\begin{enumerate}
\item
Intiliaze $p_{ij}(x)=\frac{1}{|\sa\cup0|}$ and $q_i(b)=\frac{R}{N}$, 
$\forall x, b, i, j$.
\item
Compute $v_{ji}(x)$, $\forall x, i, j$.
\item
Compute $u_{i}(b)$, $\forall b, i$.
\item
Compute $p_{ij}(x)$, $\forall x, i, j$.
\item
Compute $q_{i}(b)$, $\forall b, i$.
\item
Perform damping of messages $p_{ji}(x)$ and $q_{i}(b)$ with a damping 
factor $\Delta$.
\end{enumerate}
This completes one iteration of the LaMP algorithm. These steps are 
repeated until a fixed number of iterations are reached. 

{\em Complexity of LaMP:}
The computational complexities of evaluating the messages $v_{ji}(x)$, 
$p_{ij}(x)$, and $q_i(b)$ are of order $O(MN|\sa|)$ per iteration of the 
LaMP algorithm. Because of the $N-1$ convolutions required for each of 
the $N$ antenna activity indicators, a naive method of computing $u_i(b)$ 
will result in a computational complexity of order $O(N^3+N^2)$. To reduce 
this complexity, we propose the following techniques.

{\em 1. Deconvolution}: 
In this technique, at every iteration, we first compute 
$\phi_0=\bigotimes\limits_{i=1}^{N} q_i$, which gives a probability mass 
function with $N+1$ points $(0,1,\cdots,N)$. To evaluate each $\phi_i$, 
we deconvolve $q_i$ from $\phi_0$, which gives the required probability 
mass function with $N$ points. This method has a computational complexity 
of order $O(N^2+N)$. A disadvantage of this method is that, in a limited 
precision system, the deconvolution operation is not numerically stable. 
Because of this reason, deconvolution may some times result in negative 
values to probabilities. To avoid this limitation in fixed precision 
systems, we can employ one of the following techniques.

{\em 2. Using FFT}:
Since Fourier transform converts convolution to multiplication in the 
transformed domain, FFT can be exploited for reducing the complexity in 
computing $u_i(b)$. First, we obtain $Q_i$, an $N$-point FFT of  
$q_i, \forall i$. Then, we compute $\phi_f=\Pi_{i=1}^{N}Q_i$. Now, to 
evaluate each $\phi_i$ we compute IFFT$(\frac{\phi_f}{Q_i})$, which 
gives the required probability mass function with $N$ points. This 
method has a computational complexity of order $O(N^2+N)$. It is noted 
that the Fourier transform preserves only the total power in a signal 
(a consequence of Parseval's theorem) and not the sum of amplitudes, 
i.e., $\sum_{\forall k}Q_i(k)\neq\sum_{\forall k}q_i(k)=1$. Thus, in 
order to satisfy the condition of total probability being $1$, every 
$\phi_i$ computation requires a normalization. This marginally increases 
the computations required compared to the previous technique.

{\em 3. Gaussian approximation}:
To compute $u_i(b)$, we require $N-1$ probability mass functions, i.e., 
$q_i$s. We note that $q_i$ is a Bernoulli distribution with probability 
masses for $a_i=0$ and $a_i=1$, i.e., $\Pr(a_i=0|\vx)=q_i(0)$ and 
$\Pr(a_i=1|\vx)=q_i(1)$. The evaluation of $\phi_i$ requires the 
computation of the distribution of the random variable 
$\sum\limits_{l=1, l\neq i}^{N}a_l$, where the antenna activity 
indicator $a_i$ is distributed according to the Bernoulli distribution 
$q_i$. For large $N$, the distribution of this sum is approximately 
Gaussian with mean $m_i$ and variance $c_i$, where
\[ m_i=\E\bigg(\sum_{l=1, l\neq i}^{N}a_i\bigg)=\sum_{l=1, l\neq i}^{N}q_l(1),\]
\[ c_i=\text{Var}\bigg(\sum_{l=1, l\neq i}^{N}a_i\bigg)=\sum_{l=1, l\neq i}^{N}q_l(0)q_l(1).\]
Let $g_i\Define\sn(m_i,c_i)$. Then, $u_i(1)\propto g_i(R-1)$ and 
$u_i(0)\propto g_i(R)$. The computational complexity of this technique 
is of order $O(N)$. From simulations, we observed that the LaMP algorithm 
requires a few more number of iterations to converge when this technique 
is employed. This is because the probability masses obtain from the 
Gaussian distribution is only an approximation, while the other two 
techniques give the exact probability.

\begin{figure}
\centering
\includegraphics[width=3.5in, height=2.6in]{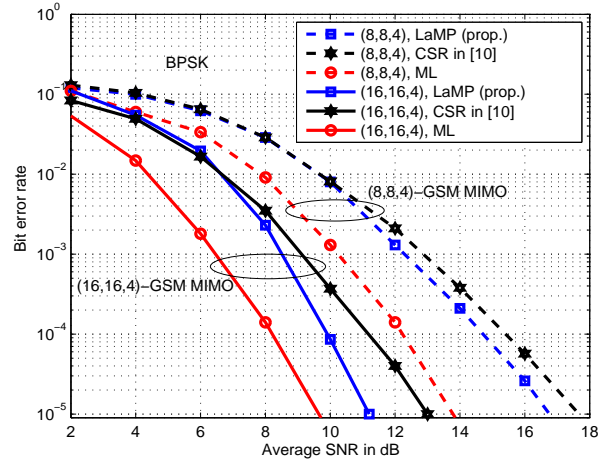}
\vspace{-4mm}
\caption{BER performance comparison between proposed LaMP detection, 
CSR detection in \cite{new_ref}, and ML detection for $(8,8,4)$- and 
$(16,16,4)$-GSM MIMO.}
\label{sim1}
\vspace{-2mm}
\end{figure}

\begin{figure}
\centering
\includegraphics[width=3.5in, height=2.6in]{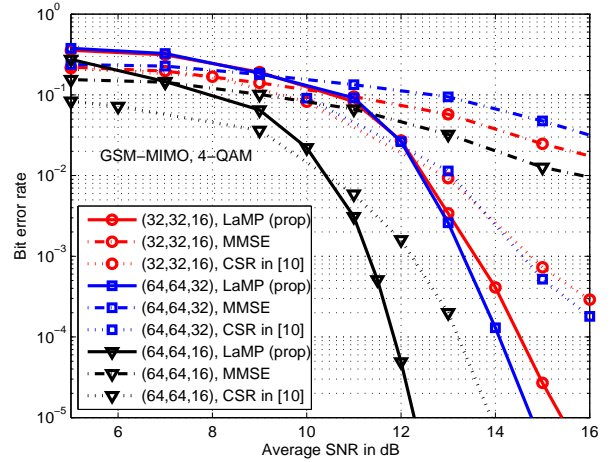}
\vspace{-4mm}
\caption{BER performance comparison between LaMP detection, CSR 
detection, and MMSE detection for $(32,32,16)$-, $(64,64,32)$- and 
$(64,64,16)$-GSM MIMO.}
\label{sim2}
\vspace{-4mm}
\end{figure}

{\em Simulation results}:
For comparison purposes, we simulated the detection performance of
convex superset relaxation (CSR) based detection in \cite{new_ref}
for GSM-MIMO. We have adapted the CSR algorithm presented for GSSK 
in \cite{new_ref} to GSM by modifying one of the CSR's constraints, 
namely, $\sum_{i=1}^{N}x_i=R$ to $\|\vx\|^2\leq R\gamma$, where 
$\gamma$ is the signal power of the modulation alphabet.

Figure \ref{sim1} presents a performance comparison between the proposed
LaMP detection, ML detection, and CSR based detection for different
$(N,M,R)$- GSM MIMO systems with BPSK. It can be seen that the LaMP
performance is away from ML performance by about 3 dB and 1.6 dB for
$(8,8,4)$- and $(16,16,4)$-GSM MIMO systems, respectively, at $10^{-5}$
BER. Thus, the performance of LaMP improves as the system dimensions 
increase. Also, LaMP detection performs better than the CSR detection in
\cite{new_ref} -- e.g., by about 1.8 dB at $10^{-5}$ BER in
$(16,16,4)$-GSM MIMO system. Figure \ref{sim2} presents a performance 
comparison between the LaMP detection, MMSE detection (performed as
$[\mh^H\mh+\frac{1}{SNR}\mi]^{-1}\mh^H\vy$), CSR detection in 
\cite{new_ref} for the following large-scale  GSM-MIMO system 
configurations: {\small ($i$) $(32,32,16)$-GSM, 4-QAM, $|\sap|=2^{29}$, 
$61$ bpcu, ($ii$) $(64,64,16)$-GSM, 4-QAM, $|\sap|=2^{48}$, $80$ bpcu, 
and ($iii$) $(64,64,32)$-GSM, 4-QAM, $|\sap|=2^{60}$, $124$ bpcu}. It 
can be seen that the proposed LaMP algorithm performs better and its
performance improves as the dimensionality of the GSM signal increases.

\begin{figure}
\centering
\includegraphics[width=3.5in, height=2.6in]{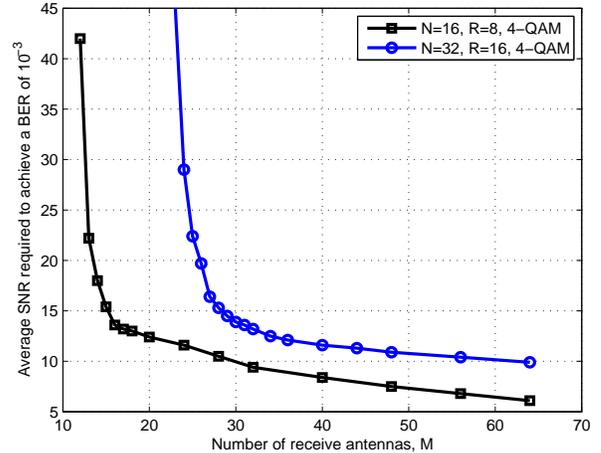}
\caption{Average SNR required by the LaMP detection algorithm to achieve a 
BER of $10^{-3}$ as the number of receive antennas $M$ is varied for 
$N=32,R=16$, and $N=16,R=8$ GSM MIMO systems with 4-QAM.}
\label{sim3}
\end{figure}

In Fig. \ref{sim3}, we plot the average SNR required by the LaMP detection
algorithm to achieve a target BER of $10^{-3}$ as the number of receive
antennas $M$ is varied. We consider two GSM MIMO systems here, ($i$)
$N=32, R=16$, 4-QAM, $61$ bpcu, and ($ii$) $N=16, R=8$, 4-QAM, $29$ bpcu.
From Fig. \ref{sim3}, it can be seen that as $M$ increases the SNR required
to achieve the target BER reduces. It can also be observed that the LaMP
detection algorithm performs well even when the system is underdetermined,
i.e., $M<N$. This is because the LaMP detection algorithm exploits the 
sparse structure of the GSM transmit signals to efficiently perform 
detection.

\section{Conclusions}
\label{sec6}
We studied GSM MIMO capacity and low complexity methods for GSM encoding 
and detection. We made the following key contributions in this paper: 
1) we derived lower and upper bounds on the GSM MIMO capacity, 2) we
proposed an efficient GSM encoding method using combinadics for large 
number of transmit antennas and RF chains, and 3) we also proposed
a message passing based low-complexity detection algorithm suited for 
large-scale GSM MIMO systems.

\end{document}